\documentclass[twocolumn,secnumarabic,superscriptaddress,showpacs,fleqn,showkeys,floatfix, amssymb, nobibnotes, aps, prb]{revtex4-2}
\usepackage{graphicx}
\usepackage{dcolumn}
\usepackage{float}
\usepackage{bm}
\usepackage{xcolor}
\usepackage{afterpage}

\begin{document}
\title{Monoclinic symmetry at the nanoscale in lead-free ferroelectric BaZr$_{x}$Ti$_{1-x}$O$_{3}$ ceramics}
\author{K. Dey}
 \address{UGC DAE Consortium for Scientific Research, University Campus, Khandwa Road, Indore 452001, India.}
 \author{A. Tripathy}
 \address{UGC DAE Consortium for Scientific Research, University Campus, Khandwa Road, Indore 452001, India.}
 \author{S. R. Sahu}
 \address{UGC DAE Consortium for Scientific Research, University Campus, Khandwa Road, Indore 452001, India.}
 \author{H. Srivastava}
 \address{Raja Ramana Centre for Advance Technology, Indore 452013, India.}
 \author{A. Sagdeo}
 \address{Raja Ramana Centre for Advance Technology, Indore 452013, India.}
 \address{Homi Bhabha National Institute, Training School Complex, Anushakti Nagar, Mumbai, 400094, India.}
 \author{J. Strempfer}
 \address{Advanced Photon Source, Argonne National Laboratory, Lemont, Illinois, 60439, USA.}
 \author{D. K. Shukla}%
 \thanks{Corresponding Author: dkshukla@csr.res.in}
 \address{UGC DAE Consortium for Scientific Research, University Campus, Khandwa Road, Indore 452001, India.}
 %\date{\today}
\begin{abstract}
Local structural symmetries play a key role in the functionalities of ferroelectric materials and are often found different from average symmetry. Here, we study the real space nanoscale structure in Pb-free BaZr$_{x}$Ti$_{1-x}$O$_{3}$ (x $\leq$ 0.10) by pair distribution function measurements, complemented by transmission electron microscopy and x-ray diffraction. Our observations show existence of the rhombohedrally distorted unit cells; however, at intermediate length scales, at least up to 5 nm, there exist nano-scale correlated regions of monoclinic symmetry. This is complemented by the observation of curved frustrated nanodomains. Further, the average structure is found to have coexisting monoclinic and rhombohedral symmetries. Our observation of a two-phase ferroelectric state is in contrast to interferroelectric instabilities of conventional polymorphic phase boundaries reported for doped BaTiO$_{3}$. 
\end{abstract}     
\maketitle

\section{Introduction}
  The BZT (BaZr$_{x}$Ti$_{1-x}$O$_{3}$) solid solution has been proven to be a model system in search of Pb-free superior piezoceramics and has been investigated intensively due to its potentially rich phase diagram by both theoreticians as well as experimentalists \cite{Akbarzadeh2012,nuzhnyy2012broadband,sherrington2013bzt,peng2018thermodynamic,mentzer2019phase}. The parent compound BaTiO$_3$ (x = 0) is a typical ferroelectric \cite{kwei1993structures}. By introducing Zr at the Ti sites, it evolves from a `conventional ferroelectric' to a `ferroelectric with diffuse phase transition' (DPT). It then turns into a `relaxor' (R) and finally becomes a `dipole glass' (DG) for a dilute concentration of Ti \cite{nuzhnyy2012broadband,petzeltbroadband}. The end compound BaZrO$_3$ is a paraelectric or `incipient ferroelectric' with simple cubic perovskite (ABO$_3$) structure \cite{akbarzadeh2005combined}. In particular, the compositions with x $\leq$ 0.1 show enhanced dielectric and piezoelectric properties \cite{maiti2011evaluation,hennings1982diffuse,dong2012enhanced,yu2000orientation,yu2002piezoelectric}. These properties have been commonly attributed to the appearance of interferroelectric phase boundaries by crystallographic structure studies through X-ray diffraction (XRD) and neutron powder diffraction (NPD) studies \cite{kalyani2013anomalous,kalyani2013polymorphic} and are generally known as polymorphic phase boundaries (PPB) \cite{garcia2020polymorphic}. In BZT ceramics, NPD revealed the coexistence of ferroelectric phases: the orthorhombic (\textit{Amm}2) phase either coexists with the tetragonal (\textit{P}4\textit{mm}) phase or the rhombohedral (\textit{R}3\textit{m}) phase for x = 0.02-0.08. A single rhombohedral (\textit{R}3\textit{m}) phase exists for x $\geq$ 0.09 \cite{kalyani2013polymorphic}. However, the average structural studies by conventional diffraction methods cannot adequately probe the local structural deviations, which exist due to the large radii mismatch between Zr$^{4+}$ and Ti$^{4+}$ (ionic radii r $\sim$ 0.72 \AA{} for Zr$^{4+}$ \& r $\sim$ 0.61 \AA{} for Ti$^{4+}$) \cite{shannon1976revised}. Further, their ferroelectric strength also has disparities \cite{Akbarzadeh2012}. These give rise to nanoscale polar order or polar nano regions (PNRs) which has been detected by diffuse electron scattering \cite{liu2007structurally,miao2009polar}, Raman spectroscopy \cite{buscaglia2014average,farhi1999raman} and through spontaneous volume ferroelectrostriction (SVFS)\cite{dey2021coexistence}. 
  
	In Pb-based solid solutions recent experimental evidences of nanoscale orders and their spatial correlations with polar domains have challenged previous theoretical interpretations \cite{takenaka2017slush,kumar2021atomic,eremenko2019local,krogstad2018relation}. In homovalent (Zr$^{4+}$, Sn$^{4+}$) substituted BaTiO$_{3}$, nanoscale correlations appear by setting up random local strain fields \cite{liu2007structurally} or by slowing down Ti$^{4+}$ dynamics \cite{pramanick2018stabilization}. This is supported also by extended X-ray absorption fine structure (EXAFS) studies, which unraveled that ZrO$_{6}$ and TiO$_{6}$ octahedra retain their individual character in BZT solid solution, similar to that in their parent compounds BaZrO$_{3}$ and BaTiO$_{3}$ \cite{laulhe2006exafs,levin2011local}. However, EXAFS is limited only to the study of the local surroundings of the absorbing atom. Understanding the exact nature of the nanoscale atomic order and their correlation with material performance in these compounds has so far remained a nontrivial undertaking. For better understandings of the relation between the local structural correlations and the average structure, and the materials performance, it requires comprehensive insight at the nanoscale level irrespective of chemical specifications, which can be gained by X-ray total scattering and pair distribution function (PDF) methodology \cite{aksel2013local,goetzee2017electric,usher2015electric}. 
 \setlength{\parskip}{0pt}

 In order to understand the role of local structural correlations behind large piezoelectric properties of BZT ceramics, we have performed x-ray PDF analysis and a set of complementary measurements including transmission electron microscopy (TEM) and XRD. These reveal a direct correlation between structural coherence weakening and polarization decoupling and their connection to monoclinicity. The presence of extended nanoscale correlated regions with monoclinic symmetry have been established from atomic PDF analysis. Mesoscale investigations by TEM unraveled hierarchical arrangements of micro-nano domains. The miniaturization of domains to the nanoscale is leading to an easy response under external influences such as electric field, which makes these BZT ceramics compatible to high piezo response. Additionally, we have observed the presence of polar nano regions in the high-temperature paraelectric cubic phase which are supposed to be a precursor of the observed nanoscale correlated regions in the ferroelectric state.    

\section{Experiment}
 For synthesis, the solid state reaction route was employed, as detailed in the Supplemental Material \cite{supp2}. X-ray total scattering measurements were carried out at room temperature at 87 keV photon energy in transmission geometry and data were collected using a Perkin Elmer area detector. Fit2D \cite{hammersley1996two} was used to extract 1D data from the 2D images. The atomic pair distribution function G(r) was obtained by processing the 1D profiles using PDFgetX3 \cite{juhas2013pdfgetx3}. The scattering contribution by the Kapton capillary used as sample container and air were subtracted. The resulting PDFs were fitted with PDFgui \cite{farrow2007pdffit2}. XRD patterns were recorded on a Bruker D2 Phaser X-ray diffractometer utilizing Cu K$_{\alpha}$ radiation. TEM measurements were performed at room temperature by utilizing the CM200 instrument operated at 200 kV. For the TEM measurements the ceramics were first thinned down mechanically to 100 $\mu$m. The specimen was then dimpled and ion milled until perforation. High temperature synchrotron X-ray diffraction (SXRD) measurements were performed at beamline BL12 at Indus-2, RRCAT (India) at $\sim$14.87 keV. Temperature dependent dielectric measurements were carried out on a home developed setup utilizing HIOKI LCR meter (IM3536) and CRYOCON (22C) temperature controller. Temperature dependent \textit{d}$_{33}$ measurements were carried out by utilizing a commercial \textit{d}$_{33}$ meter (American piezo ceramics) and a miniature heating stage. Prior to XRD, X-ray PDF and TEM measurements, samples were annealed at 350$^{\circ}$C for 2 hrs to relieve mechanical stress.      
\begin{figure}[b]
\includegraphics[width=\linewidth]{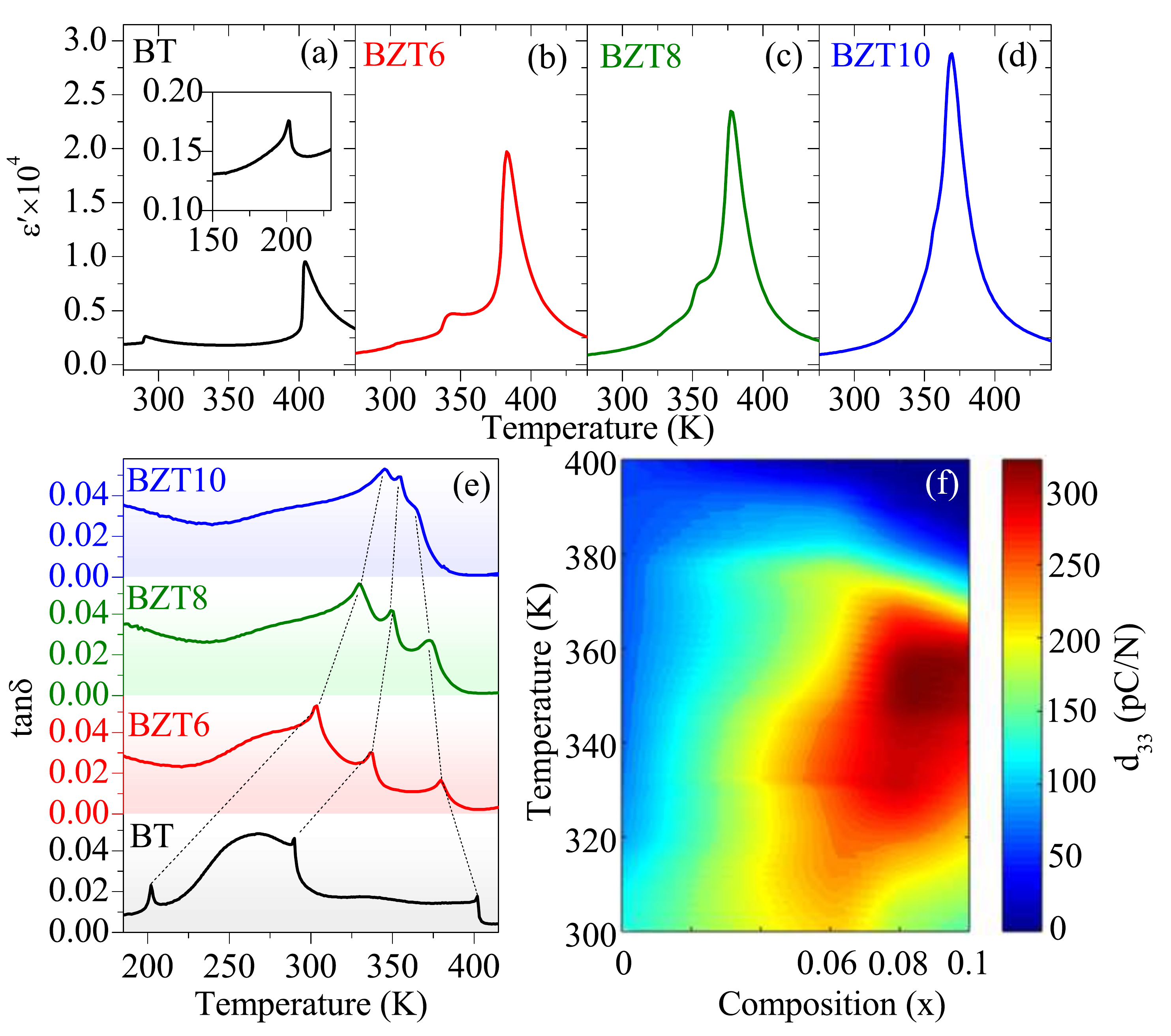}
\caption{\label{fig:epsart} Temperature dependence of real permittivity of BaZr$_{x}$Ti$_{1-x}$O$_3$ (a) BT (x = 0.0), (b) BZT6 (x = 0.06), (c) BZT8 (x = 0.08), and (d) BZT10 (x = 0.10) at 1 kHz during heating cycle. All graphs are plotted on the same x-y scale for better visualization of relative changes. The inset of (a) shows the low temperature anomaly in permittivity for BaTiO$_{3}$. (e) Temperature-dependent dielectric loss (tan$\delta$) at 1 kHz frequency for the above four compositions during the heating cycle. (f) Temperature composition contour plot of \textit{d}$_{33}$.}
\label{Comp_die_piezo}
\end{figure}

%\begin{figure}
%\includegraphics[width=\linewidth]{BZT10_Dielectric2.eps}
%\caption{\label{fig:epsart} Temperature dependence of (a) real \& (b) imaginary parts of relative permittivity at different frequencies. The inset in (a) shows the zoomed portion of real permittivity maxima. (c) Curie-Weiss fit of the real permittivity data shows a deviation from the temperature marked with arrow ($\sim$ 450 K) \& the inset in (c) shows the linear fitting by using the empirical formula (see text). (d) Temperature evolution of cubic lattice parameter. Inset shows the temperature variation of calculated spontaneous volume ferroelectrostriction (SVFS).} 
%\label{BZT10_die}
%\end{figure}
\setlength{\parskip}{0pt}

\begin{figure}
\includegraphics[scale=0.25]{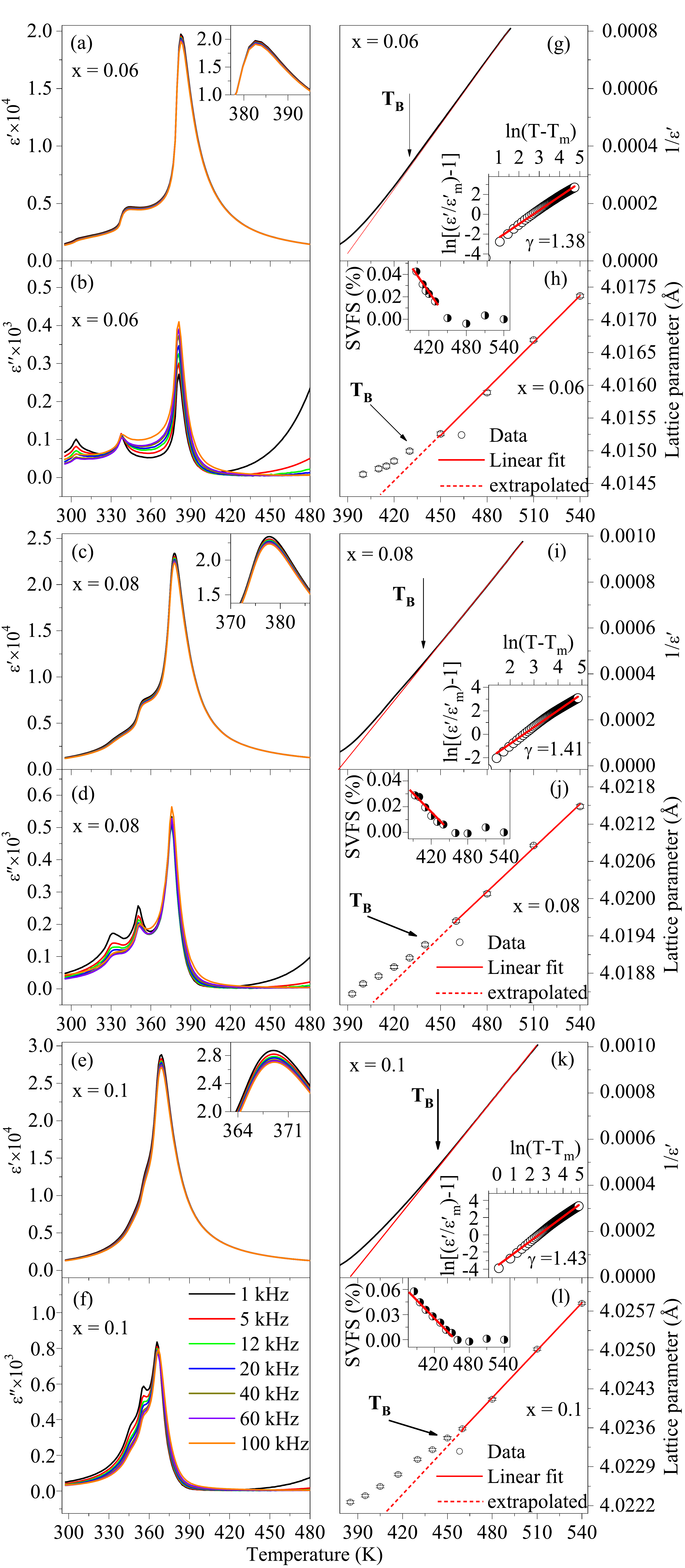}
\caption{\label{fig:epsart} Temperature dependence of real [(a), (c), (e)] and imaginary [(b), (d), (f)] parts of relative permittivity at different frequencies for BZT6 (x = 0.06), BZT8 (x = 0.08), and BZT10 (x = 0.10), respectively. The insets in panels [(a), (c) and (e)] show the enlarged portion of real permittivity maxima. [(g), (i), (k)] Curie-Weiss fit of the real permittivity data shows a deviation from the temperature marked as T$_{B}$ with arrow and the insets in panels (g), (i) and (k) show the linear fitting by using the empirical formula (see text). [(h), (j), (l)] Temperature evolution of cubic lattice parameters from SXRD data. Insets in panels (h), (j) and (l) show the temperature variation of calculated spontaneous volume ferroelectrostriction (SVFS).}
\label{Die_all}
\end{figure} 
 
\section{Results and discussions}
 For BaTiO$_{3}$ (BT) three sharp relative permittivity peaks are observed at around 402 K, 290 K and 202 K (see Fig.~\ref{Comp_die_piezo} (a) \& (e)). With decreasing temperature, the cubic \textit{Pm}$\bar{3}$\textit{m} phase of BaTiO$_3$ transforms into the tetragonal \textit{P}4\textit{mm} phase at 402 K, which on further cooling transforms into the orthorhombic 
 \textit{Amm}2 phase at 290 K and finally becomes rhombohedral \textit{R}3\textit{m} at 202 K \cite{fu2015role,kwei1993structures}. BaZr$_{x}$Ti$_{1-x}$O$_3$ (x $\leq$ 0.10) ceramics show close similarity in the sequence of above thermodynamic phase transitions of the parent compound BaTiO$_3$ (see Fig.~\ref{Comp_die_piezo} (b-d) \& (e)) which is consistent with earlier reports \cite{kalyani2013polymorphic}. The anomalies at lower temperatures for higher Zr doped compositions are looking like sluggish humps in $\varepsilon^{\prime}$ vs. T. They become more evident in the temperature dependent dielectric loss (tan$\delta$) spectra. Fig.~\ref{Comp_die_piezo} (f) shows the temperature dependence of piezoelectric charge coefficient (\textit{d}$_{33}$) for all these compositions. At room temperature the Zr doped compositions exhibit enhanced \textit{d}$_{33}$ with respect to the parent compound. Interestingly, the BZT10 composition shows increasing \textit{d}$_{33}$ with temperature and which becomes maximum ($\sim$ 316 pC/N) at an elevated temperature of about 355 K. 
  
\begin{figure}
\includegraphics[width=\linewidth]{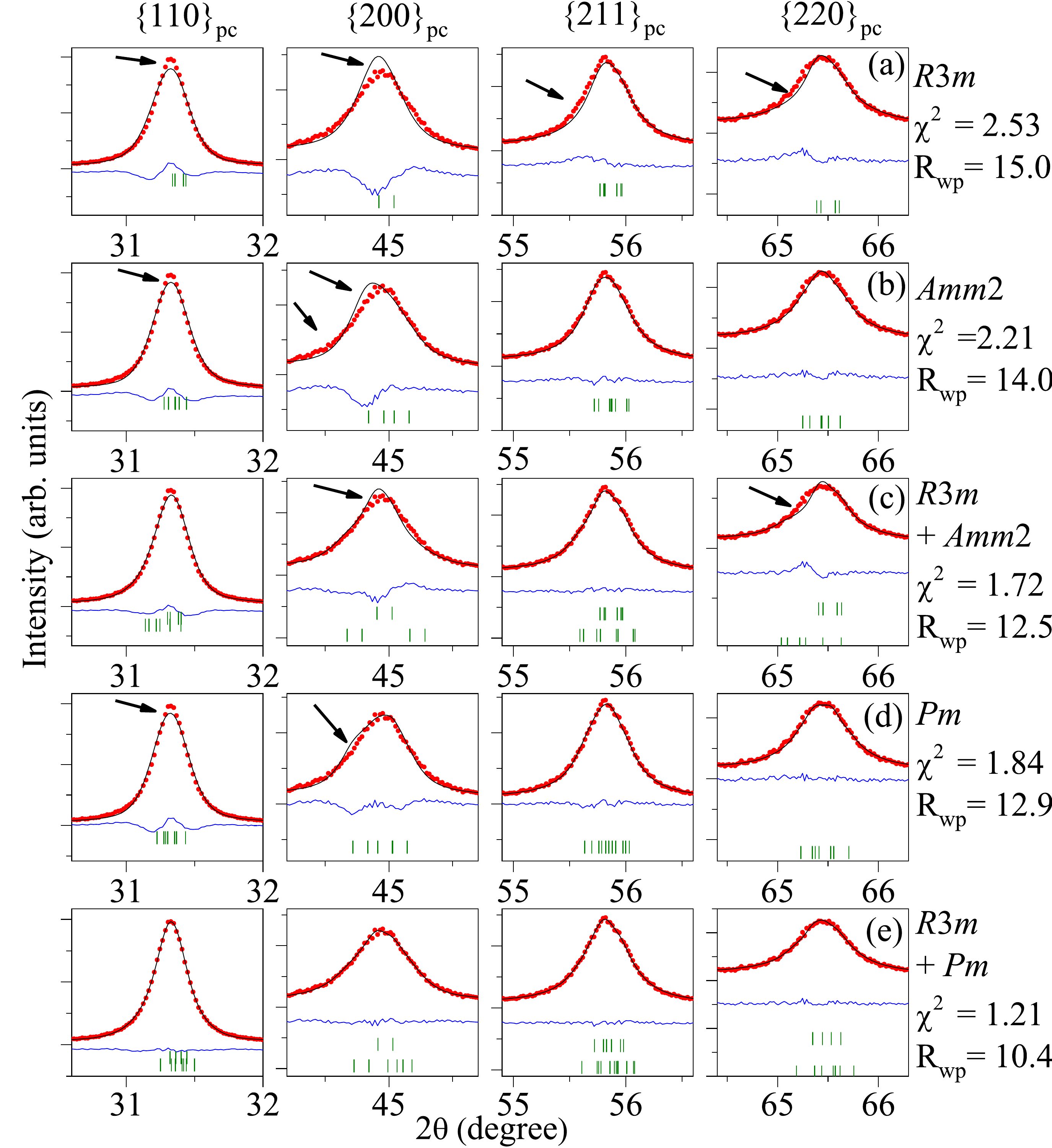}
\caption{\label{fig:epsart} (a) Rietveld refined XRD profiles of selected reflections in BZT10 using (a) \textit{R}3\textit{m}, (b) \textit{Amm}2, (c) \textit{R}3\textit{m} + \textit{Amm}2 (d) P\textit{m}, (e) \textit{R}3\textit{m} + \textit{Pm} models. Arrows indicate the misfit regions.}
\label{XRD_BZT10}
\end{figure} 

 \begin{figure}
\includegraphics[width=\linewidth]{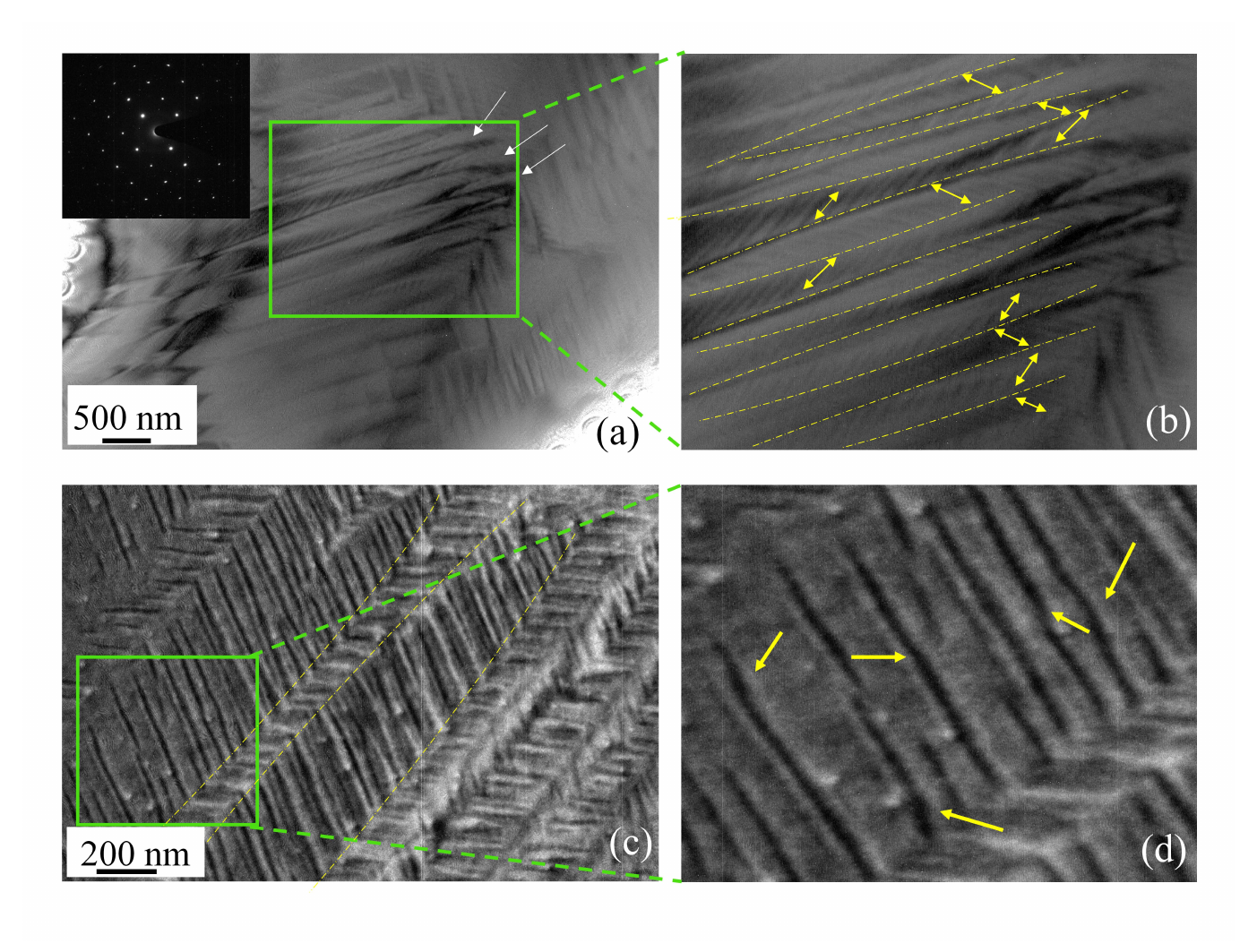}
\caption{\label{fig:epsart} (a) Room temperature domain morphology of BZT10 as seen by TEM. The white arrows indicate the wedge shape of microdomains. The inset shows the selected area electron diffraction (SAED) pattern. (b) An enlarged view of the selected micrograph in panel (a) for better visualization of nanoscale domains inside microdomains. The dash-dotted curve indicates microdomain walls. Bidirectional yellow arrows indicate the directions of extension of nanodomains. (c) The nanoscale domain pattern of the same sample at smaller length scale. (d) The enlarged view of selected region in panel (c) shows striation-like nano domains. The yellow arrows show the bend locations.}
\label{TEM_BZT10}
\end{figure}
 
 Fig. \ref{Die_all} (a-f) show the temperature dependent  $\varepsilon^{\prime}$ and $\varepsilon^{\prime\prime}$, measured at selected frequencies for BZT6, BZT8, BZT10 samples. The three frequency-independent peaks in $\varepsilon^{\prime\prime}$ vs. temperature is in accordance with the three structural transitions in these compositions \cite{kwei1993structures}. The Zr doping in BaTiO$_{3}$ is known to stabilize the rhombohedral structure \cite{buscaglia2014average}. Our dielectric data also show that above room temperature Zr-doped samples have already passed through well-known three phase transitions of BaTiO$_{3}$ \cite{kwei1993structures}, and these appeared to possess rhombohedral structure at room temperature. Insignificant dielectric dispersion is observed for all the compositions at around dielectric maxima temperature (T$_{m}$) (insets of Fig. \ref{Die_all} (a), (c) \& (e)). Moreover, shifting of the permittivity peaks to higher temperatures with higher frequencies are within 1 K temperature. A noticeable departure from Curie-Weiss behavior was found below $\sim$ 430 K for BZT6 (see Fig. \ref{Die_all} (g)), below $\sim$ 440 K for BZT8 (see Fig. \ref{Die_all} (i)) \& below $\sim$ 450 K for BZT10 (see Fig. \ref{Die_all} (k)) which indicate characteristic Burn's temperature, T$_{B}$ \cite{burns1983glassy,bokov2006recent}. Burn's temperature usually characterizes the onset of polar nano regions (PNRs) in a diffused ferroelectric \cite{chen2013effectively}. To estimate the degree of diffusion, we have calculated the diffusion exponent $\gamma$ by fitting the permittivity data with an empirical formula\cite{bokov2003empirical}, $\frac{\varepsilon_{m}}{\varepsilon}=1+\frac{(T-T_{m})^\gamma}{2\delta^{2}}$, where $\varepsilon_{m}$ is the real permittivity maxima at a temperature T$_{m}$. The calculated values of the $\gamma$ varies between 1.38-1.43 (for these compositions) which characterizes these as a diffuse ferroelectric phase transition, an intermediate between classical ferroelectric ($\gamma$=1) and relaxor ($\gamma$=2) \cite{nuzhnyy2012broadband}. This is analogically similar to PMN-PT, PZN-PT compositions nearby MPB \cite{noblanc1996structural,li2016origin}.
 To further validate the presence of the PNRs above T$_m$ and to have insights of the characteristic temperature T$_{B}$ observed in dielectric measurements, we now focus on the results of temperature-dependent structural studies obtained through SXRD measurements. Interestingly, below (T$_{B}$) cubic lattice parameter deviates from linear behavior (Fig. \ref{Die_all} (h), (j) \& (l)). Such deviation without a phase transition is not normal and in present case the deviation marks the onset of PNRs. The anharmonic lattice phonon vibration decreases by lowering temperature and local ferroelectric order in PNRs starts to overcome this and results in dilation of lattice parameters. This effect is termed as spontaneous volume ferroelectrostriction (SVFS) \cite{chen2015negative}. We have calculated SVFS by using the equation $\omega$$_s$= (V$_{exp}$-V$_{nm}$)/V$_{nm}$$\times$100$\%$ where V$_{exp}$ is the observed volume and V$_{nm}$ is the usual behavior of volume \cite{chen2013effectively,dey2021coexistence} shown in the insets of Fig. \ref{Die_all} (h), (j) \& (l). The evolution of SVFS just below T$_B$ confirms existence of the PNRs in these.
 
\begin{figure}[b]
\includegraphics[width=\linewidth]{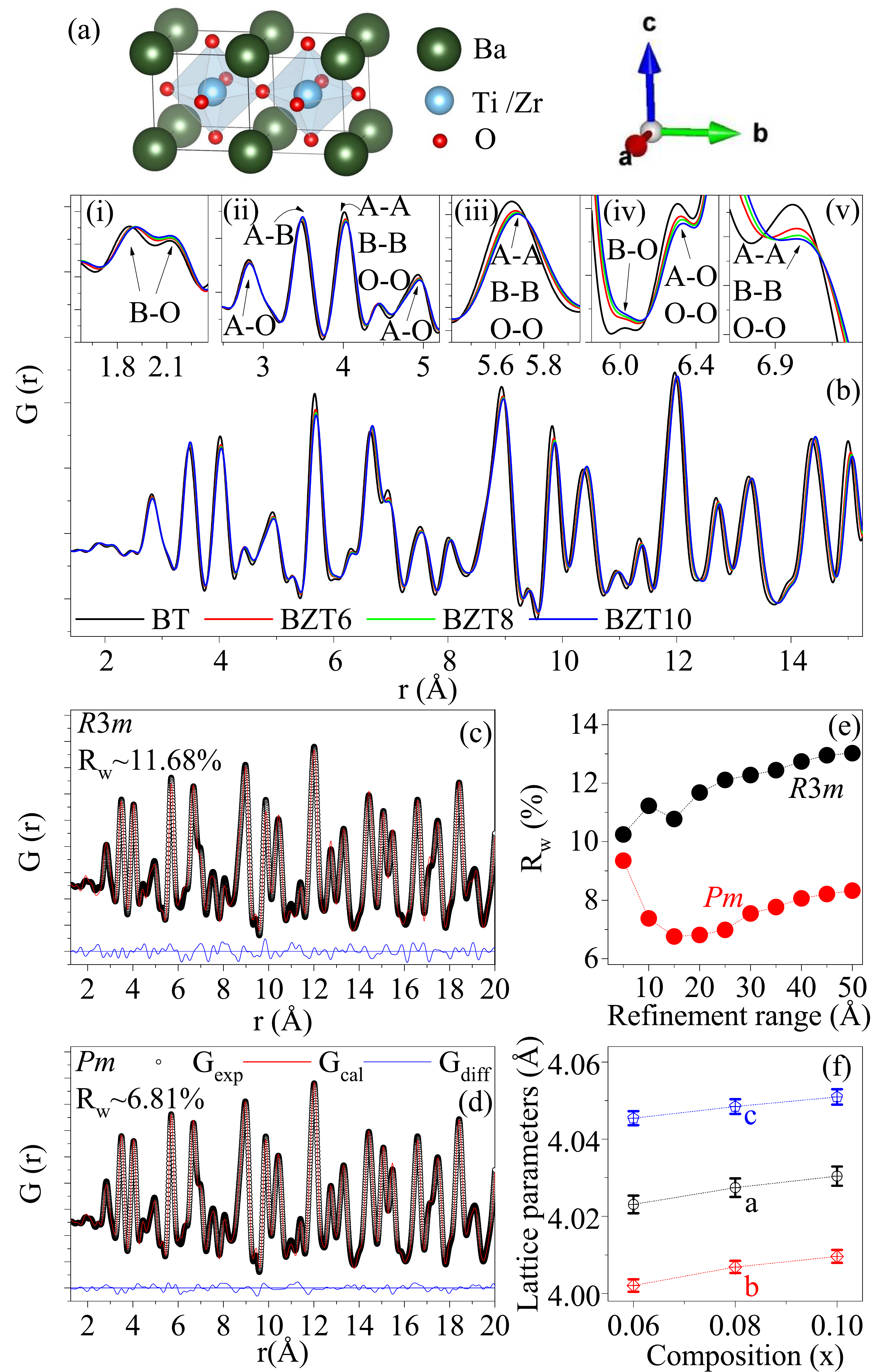}
\caption{\label{fig:epsart} (a) Schematic diagram of perovskite ABO$_{3}$ structure. (b) Experimental pair distribution functions (PDFs) of BT, BZT6, BZT8, and BZT10 in the small \textit{r} regime (up to 15.25 \AA) at room temperature. The insets (i)-(v) represent selected peaks which correspond to the atomic pairs shown by arrows, where A denotes Ba and B denotes Ti/Zr atoms. The experimental PDFs, calculated PDFs and their difference obtained by real space structure refinement of BZT10 up to 20 \AA{} using (c) \textit{R}3\textit{m} and (d) \textit{Pm} models. (e) The agreement factor R$_{w}$ fitted with \textit{R}3\textit{m} and \textit{Pm} models as a function of refinement range. (f) The Zr concentration (x) dependence of lattice parameters.}
\label{PDF1}
\end{figure}
 %(f) shows the evolution of lattice parameters \textit{a, b, c} \& (g) inter-axial angle (\textit{$\beta$}) obtained by fitting with different ranges using Pm structural model for BZT10.  Similar fitting parameters for BZT6 and BZT8 are presented in the supplements  \cite{supp2}.
\begin{figure}
\includegraphics[width=\linewidth]{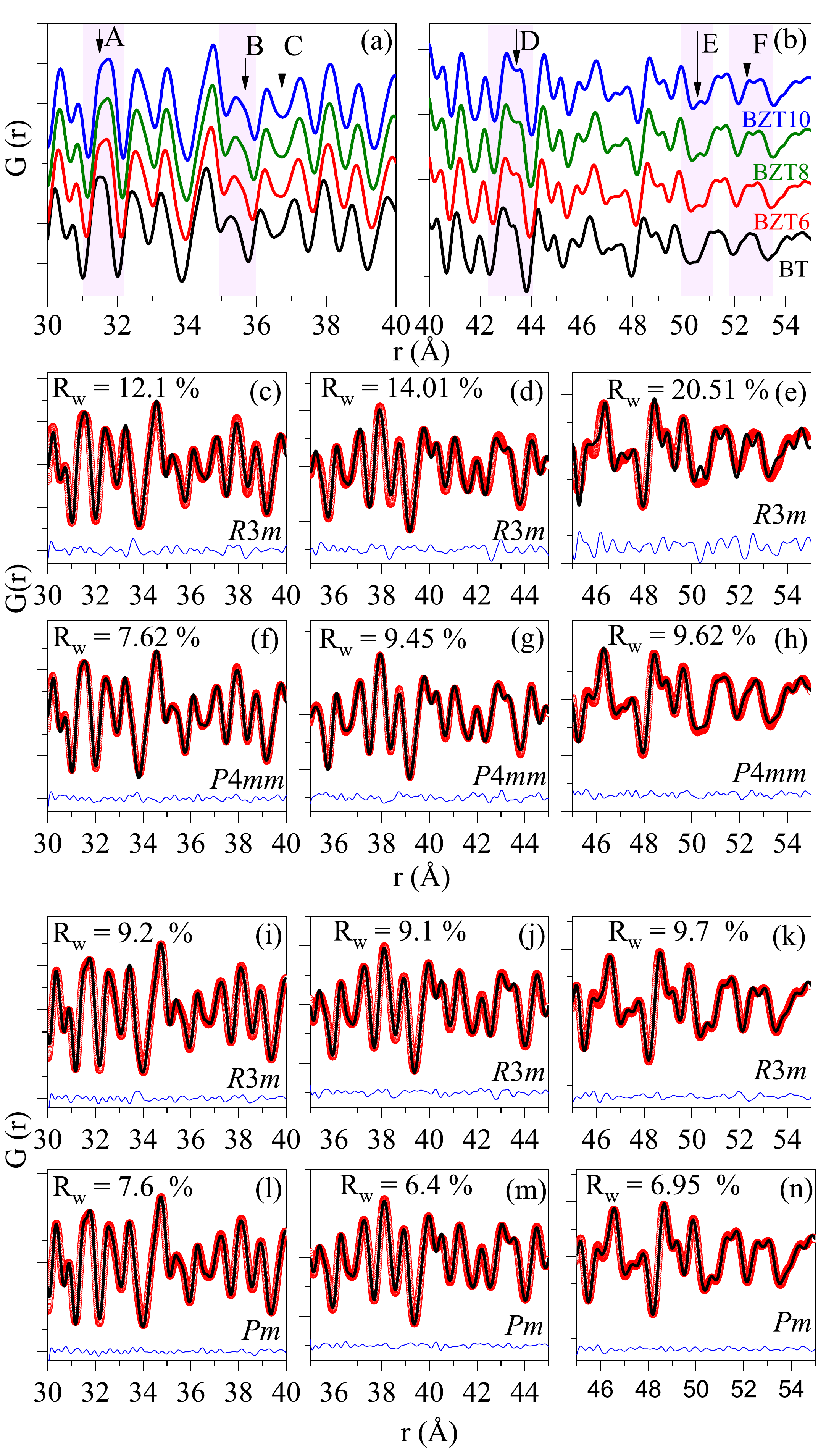}
\caption{\label{fig:epsart} [(a) \& (b)] Experimental PDF pattern in intermediate to larger \textit{r} ranges for BT, BZT6, BZT8, and BZT10. Fitted PDF patterns of the BT by using \textit{R}3\textit{m} [(c), (d), (e)] and by using \textit{P}4\textit{mm} [(f), (g), (h)]. Fitted PDF patterns of the BZT10 by using \textit{R}3\textit{m} [(i), (j), (k)] and by using \textit{Pm} [(l), (m), (n)}
\label{Box_car_higher}
\end{figure}

 In general, the emergence of PNRs from much above T$_m$ have serious influences on the structure in the ferroelectric phases \cite{brajesh2015relaxor, li2016origin}. For the detailed structural analysis we have selected BZT10, which is supposed to posses rhombohedral structure at room temperature \cite{buscaglia2014average}. Line shapes of \{110\}$_{pc}$, \{200\}$_{pc}$, \{211\}$_{pc}$ \& \{220\}$_{pc}$ for BZT10 are shown in Fig. \ref{XRD_BZT10}. The rhombohedral \textit{R}3\textit{m} model cannot account the broad line shape of \{200\}$_{pc}$ and asymmetries in line shape for \{211\}$_{pc}$ \& \{220\}$_{pc}$. The asymmetries are dominant at the left shoulders for these reflections and are partly accompanied with diffuse scattering. Rietveld analysis with \textit{Amm}2 model can not achieve the satisfactory fit either. A combination of these two also cannot account fitting discrepancies. One, two or three phase possible symmetry combinations of BT (\textit{R}3\textit{m}, \textit{Amm}2 and \textit{P}4\textit{mm}) are also tested and found unsatisfactory. Next, refinement have been carried out using further possible lower symmetries. A monoclinic (\textit{Pm}) structural model is found to account those asymmetric shoulders with better agreement factor than other single phase models (for detail see TABLE S1 in \cite{supp2}). However, noticeable discrepancies are observed which suggests for the requirement of an additional space group (S.G.) for accounting the whole Bragg profile.
The best fit is achieved only when monoclinic (\textit{Pm}) symmetry is combined with \textit{R}3\textit{m} (see full profile XRD patterns \& detailed fitting parameters in the Supplemental Material \cite{supp2}). 

 Our XRD results indicated the presence of the low symmetry monoclinic phase that may have strong influence on the mesoscale domain structure in these compounds. To visualize the domain structure, we next carried out TEM measurements on BZT10 which has maximum monoclinic phase fraction (see Tables S2 and S3 in the Supplemental Material \cite{supp2}). Figures \ref{TEM_BZT10} (a-d) show TEM micrographs of BZT10. Wedge-shaped microdomains are observed, as shown by white arrows (Fig. \ref{TEM_BZT10} (a)), which is typical for rhombohedral crystal structure. However, these microdomains exhibit curved walls shown by yellow dashed lines ( Fig.~\ref{TEM_BZT10} (b)). Inside those wedge-shaped domains, different sets of nanodomains exist. These colonies of high density nanodomains inside the curved walls can be visualized better in the micrograph taken at high image resolution (Fig. \ref{TEM_BZT10} (c)). The enlarged micrograph of Fig. \ref{TEM_BZT10} (c) [Fig. \ref{TEM_BZT10} (d)] shows the embedded nanodomains are not like straight pillars, but instead have irregular curved striation-like features. This kind of frustrated domains usually appears around morphotropic phase boundaries (MPB) due to the continuous decoupling of polarization direction with respect to the underlying crystal lattice \cite{khachaturyan2010ferroelectric}. Therefore, hierarchical domain arrangements appear and higher piezo-response has been found to be typically always associated with such hierarchical domain configurations \cite{hu2020ultra,wang2006hierarchical}.

 From the average structure and microstructure studies of the BZT10 a connection between enhanced functional properties and nano-scale correlations along with complex domain configurations can be readily seen. To better understand the local structural deviations on shorter length scales we have utilized X-ray total scattering and PDF analysis \cite{egami2003underneath}. Representative PDFs for the different BZT compositions along with that of undoped BaTiO$_3$ are shown in Fig. \ref{PDF1} (b) for, up to 15.25 \AA. For all compositions, the PDF peaks for Ti/Zr-O bonds split into two with almost equal intensity at $\sim$ 2 \AA{} (see inset (i) in Fig. \ref{PDF1} (b)) which represents a rhombohedral like distortion of B-O bonds and is in agreement with previous findings in BaTiO$_{3}$ \cite{senn2016emergence,page2008local,culbertson2020neutron}. Several qualitative differences can be seen in the experimental PDFs for the nearest and the next nearest atomic neighbors with increase in the Zr concentration (see insets (ii), (iii), (iv) \& (v)). Substantial changes in the PDF peak height and the width can be observed at $\sim$ 4 \AA, $\sim$ 5.6 \AA{} and $\sim$ 6.8 \AA i. e. whenever the atoms meet with each other (e.g. at \textit{r} = \textit{a}, $\sqrt{2}$\textit{a}, $\sqrt{3}$\textit{a}, where \textit{a} is the lattice parameter of the pseudocubic ABO$_{3}$ structure). This clearly indicates a change in the degree of order of the underlying atom pair correlations. The PDF peaks related to next neighbor B-O bonds at $\sim$ 6 \AA{} (inset (iv)) become much broader for the doped compound compared to the undoped, suggesting a static disorder between Ti/Zr \& next neighbor oxygens. These observations directly indicate the weakening of structural coherence among the nearest neighbors with increasing Zr concentration.
  
 In order to extract quantitative information about the pair distributions experimental PDF patterns have been fitted. The pattern fitted with the most likely symmetry (\textit{R}3\textit{m}) turned out to be highly unsatisfactory (Fig. \ref{PDF1} (c), x = 0.1) with R$_w$ = 11.68\%. All the experimental PDF peaks are accounted well with monoclinic \textit{Pm} symmetry (see Fig. \ref{PDF1} (d)) resulting in minimum R$_w$ ($\sim$6.81\%). The R$_w$ factors for variable \textit{r}-range (1.1 to r$_{max}$) fittings \cite{qiu2005orbital, culbertson2020neutron} up to r$_{max}$ 50 \AA{} with \textit{R}3\textit{m} and \textit{Pm} models indicate that the \textit{Pm} structural model accounts the experimental PDF patterns better than \textit{R}3\textit{m} (see Fig. \ref{PDF1} (e)). For the other two compositions also the experimental G(r) are fitted well with \textit{Pm} model (see Supplemental Material \cite{supp2}). The evolution of the lattice parameters with composition and their increasing behavior is in accordance with the larger ionic radii of the substituted Zr ion (see Fig. \ref{PDF1} (f)). From the variable r-range refinement method (1.1 to r$_{max}$) the fitting agreement factor has no tendency to converge towards the long-range rhombohedral phase as predicted from earlier phase diagrams \cite{kalyani2013polymorphic, maiti2008structure}. At local length scale (up to $\sim$ 5 \AA) both the models (\textit{R}3\textit{m} \& \textit{Pm}) can define the PDF peaks with almost same fitting agreement. A good match with the \textit{R}3\textit{m} model at this length scale is due to the presence of rhombohedrally distorted unit cells, similar to that of BaTiO$_{3}$ \cite{senn2016emergence}. However, for higher r-range (above $\sim$ 10 \AA) the calculated PDF pattern generated by \textit{R}3\textit{m} model starts to deviate from the experimental atomic PDFs and the discrepancies are larger for intermediate r-range. Whereas the \textit{Pm} model provides better agreement between the calculated pattern and experimental pattern over length-scales greater than $\sim$ 10 \AA{} suggesting that the intermediate range structure is \textit{Pm}. 
 
 We have also carried out fittings by taking fixed length boxes and by varying the r range sequentially, which is known as thee box-car refinement approach \cite{usher2016local, kong2021local}. This approach is suited better for estimate of length scales of local atomic order. Fig. \ref{Box_car_higher} shows the enlarged experimental PDFs of BT, BZT6, BZT8 and BZT10 as well as fitted PDFs of BT (Fig. \ref{Box_car_higher}(c-h)) and BZT10 ( Fig. \ref{Box_car_higher}(i-n)) for comparisions of evolutions of pair correlations and suitability of structural models at higher length scales. From the visual inspection of Figs. \ref{Box_car_higher} (a) and \ref{Box_car_higher} (b), the changes in the PDF patterns of the Zr-doped compositions at higher interatomic distances (medium- or long-range structure) starts to become greater with respect to that of undoped. The major changes are shown by arrows in the shaded regions. However, with increasing the Zr doping concentration (6\%  to 10\% ), no new feature appears in the PDF rather the overall PDF systematically shifts toward higher r. That means, qualitatively, medium to long-range structures of the Zr-doped compositions are similar, but different from the parent BaTiO$_{3}$. A quantitative structural refinement of the PDF patterns have been carried out for parent BT and for BZT10 by box-car approach. Our analysis clearly shows that for BT the \textit{R}3\textit{m} model (very local pair correlation) cannot account better than the \textit{P}4\textit{mm} model at higher length scales. In the same fashion comparison of \textit{R}3\textit{m} and \textit{Pm} models for the BZT10 clearly shows that \textit{R}3\textit{m} model may be accepted only at a very local scale; however, it cannot describe the pair correlations better than \textit{Pm} at intermediate scales (Fig. \ref{Box_car_higher}(i-n)). The detailed refined patterns covering low to high r range (up to 30 \AA) using the box-car approach are shown in the Supplemental Material \cite{supp2}. Finally, PDF analysis suggests the presence of the monoclinic (\textit{Pm}) phase in the intermediate \textit{r}-range (beyond nearest neighbor environments).

 We have studied the atomistic insights of BZT ($\leq$ 0.1) ceramics from local, intermediate, and average length scales. The average structure cannot adequately be described by Rietveld analysis considering the conventional \textit{R}3\textit{m} model. Also, the inclusion of the \textit{Amm}2 symmetry cannot improve the quality of the fit. This rules out any inter-ferroelectric like instability as it is commonly considered in conventional polymorphic phase boundaries (PPBs) \cite{kalyani2013polymorphic,garcia2020polymorphic,kalyani2015polarization}. Instead, we propose the higher piezoresponse to arises due to the presence of nanoscale correlations, in agreement with our microscopic results. The nanoscale correlated regions are supposed to be composed of strained TiO$_6$ octahedra surrounded by stress centers, i.e., larger ZrO$_6$ octahedra. The virtue of the presence of strained polyhedrons are detected also at higher temperature (well above T$_{m}$) in the form of PNRs. The deviation from the Curie-Weiss law and the dilation of lattice parameters from much above T$_{m}$ in these ceramics confirm that they enter into the ferroelectric phase via a diffuse phase transition and exhibit PNRs above T$_m$. Our PDF analysis confirmed the presence of monoclinic symmetry in the intermediate length scales. The intermediate structure suggests that these nanoscale correlated regions can be described better with a low-symmetry monoclinic (\textit{Pm}) space group and the \textit{Pm} symmetry is found to persist well above the intermediate length scale (from combined higher \textit{r}-range PDF analysis \& XRD analysis by Rietveld refinement). The inhomogeneous strain distribution due to the difference in structures at different length scales disrupts long-range order and consequently nanoscale domains are formed, which are clearly seen in our TEM micro-graphs. The frustrated nanodomains inside the wedge-shaped domains appear due to continuous polarization decoupling \cite{khachaturyan2010ferroelectric}. The complicated mesoscale assembly of micro-nano domains and their hierarchical arrangement is compatible with high piezoresponse in these ceramics. Theoretical investigations also confirmed the stabilization of monoclinic phase in the form of hierarchical domain arrangements when long range elastic and electrostatic interactions are incorporated with the sixth order Landau free energy polynomial \cite{ke2013formation, ke2020polarization}.    

\section{Conclusions}
In conclusion, we have studied the atomistic picture of the high piezoresponse in Pb-free BaZr$_{x}$Ti$_{1-x}$O$_3$ (x$\leq$0.1). Our PDF analysis shows that the local atomic pair correlations below $\sim$ 3.5 \AA{} appear to possess rhombohedral distortions and are essentially identical for all the compositions irrespective of their average structures. Above 3.5 \AA, the atom pair correlations start to weaken with the Zr doping. Correlation weakening by the Zr-substitution ultimately results in monoclinic distortions, which are found to be present in the intermediate range and beyond (up to at least $\sim$ 50 \AA) and are evidenced in the different \textit{r} range PDF fittings as well as in the box-car fitting. Average structural studies by XRD substantiates to the presence of the monoclinic (\textit{Pm}) phase observed in the PDF analysis. Our study of the average structure confirms the coexistence of monoclinic and rhombohedral symmetries. The observation of two phase ferroelectric state presented here is not a manifestation of interferroelectric instabilities reported as conventional PPB in the BaTiO$_{3}$ based compounds. Further, below the critical temperature T$_{B}$, the presence of the polar nano regions are confirmed through SVFS. The nanoscale polar entities observed at higher temperatures are the precursor states of the ferroelectric phase below T$_m$ and the average structural evolution is supposed to be greatly influenced by correlations within the nanoscale polar order. The mesoscale domain configurations observed by TEM measurements reveals the presence of colonies of nanodomains within the wedge-shaped macrodomains and their hierarchical arrangement. The hierarchical assembly of micro-nano domains and their greater flexibility with external perturbations results in the higher piezo-response.          

\section{Acknowledgments}
 Authors acknowledge M. N. Singh for help during temperature dependent XRD measurements and Partha Sarathi Padhi for help during sample preparation for TEM measurements. This research used resources of the Advanced Photon Source, a U.S. Department of Energy (DOE) Office of Science User Facility at Argonne National Laboratory and is based on research supported by the U.S. DOE Office of Science-Basic Energy Sciences, under Contract No. DE-AC02-06CH11357. The mail-in program at Beamline 11-ID-B contributed to the data.

\bibliography{MD_BZT}

\end{document}